\documentclass[twocolumn,showpacs,preprintnumbers,prb,fleqn]{revtex4}
\usepackage{epsfig}
\usepackage{graphicx}
\usepackage{amsfonts}

\newcommand{\ct}{\cite}
\newcommand{\bi}{\bibitem}
\newcommand{\be}{\begin{equation}}
\newcommand{\ee}{\end{equation}}
\newcommand{\ba}{\begin{eqnarray}}
\newcommand{\ea}{\end{eqnarray}}

\begin{document}
                                                                                

\title{Effect of long range connections on an infinite randomness fixed point
associated with the quantum phase transitions in a transverse Ising model}
\author{Amit Dutta}
\email{dutta@iitk.ac.in}
\author{Loganayagam, R}
\affiliation{Department of Physics, Indian Institute of Technology Kanpur - 208016, India}
\date{\today}
\begin{abstract}

We study the effect of long-range connections on the infinite-randomness fixed
point associated with the quantum phase transitions in a transverse Ising
model (TIM). The TIM resides on a long-range connected lattice where any 
two sites at a distance $r$ are connected with  a non-random ferromagnetic
bond with a probability
that falls algebraically with the distance between the sites as  
$1/r^{d+\sigma}$. The interplay of the fluctuations
due to dilutions together with the quantum fluctuations due to the 
transverse field 
leads to an interesting critical behaviour. The exponents
at the critical fixed point (which is an infinite randomness 
fixed point (IRFP)) 
are related to the classical ``long-range" percolation exponents. 
The most interesting
observation is that the gap exponent $\psi$ is exactly
obtained for all values of $\sigma$ and $d$. Exponents 
depend on the range parameter $\sigma$ and show a crossover to short-range
values 
when  $\sigma \geq 2 -\eta_{SR}$ where $\eta_{SR}$ is the anomalous dimension
for the conventional percolation problem.
Long-range connections are also found to tune the 
strength of the Griffiths phase. 
\end{abstract}
\maketitle

The presence of quenched randomness drastically modifies the nature of 
the zero-temperature (quantum) phase transitions 
\ct{fisher92,sachdev99,dutta96}. 
In contrary to the pure
systems, many low-energy and low frequency properties of systems with quenched
randomness are dictated by locally ordered ``rare regions" \ct{vojta06}. 
These active rare
regions produce stronger effects on zero-temperature transitions 
than on finite-temperature classical transitions \ct{fisher92,fisher99}. 

 Some of the novel features associated 
with low-dimensional random quantum transitions  
include activated (quantum) dynamical scaling at the quantum 
critical point and 
the existence of  Griffiths-McCoy (GM) singular regions 
\ct{griffiths69, mccoy69}
where the response function diverges even away from the critical point.
The existence of the above mentioned features  well-established
in the case of  one dimensional random  quantum Ising models
 by using a  novel  renormalisation group (RG) technique \ct{dasgupta80}
 which is exact in 
the asymptotic limit \ct{fisher92}. For $d=1$, the RG flow on the
the critical manifold for strong 
randomness  is towards an 
infinite randomness fixed point where the quenched randomness 
effectively grows stronger and stronger as the system is 
coarse-grained \ct{fisher92}. 
Numerical studies using an extension of the above RG scheme \ct{motrunich00} 
 as well as quantum Monte Carlo studies \ct{pich98} predict a similar 
scenario of IRFP  even for random quantum Ising transitions for
spatial dimension $d=2$.
Also
at an IRFP, stronger couplings always dominate and hence
 frustration turns out to be irrelevant.  Therefore, the critical behaviour
of  quantum Ising spin-glass and random ferromagnetic models 
are expected to be governed by the same fixed point \ct{fisher99}. 
Long-range spatial  correlations of disorder are found to 
enhance the off-critical singularities \ct{rieger99}. 

Situation becomes very interesting when the quenched randomness arises due to
dilution or vacancies in a quantum magnetic model \ct{harris74, senthil96, vojta05, sandvik02,dutta03}.
The problem is  immediately connected to 
the well-studied problem of 
the percolation transition of the underlying lattice \ct{stauffer92}.
The study of quantum phase transitions
in a TIM  at the percolation threshold of a dilute lattice shows
the existence of
 an infinite randomness critical 
point scenario in dimensions greater than unity \ct{senthil96}.  

\begin{figure}
\includegraphics[height=1.3in,width=2.1in]{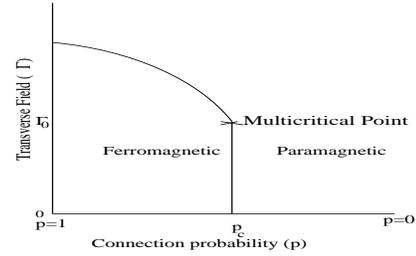}
\caption{ The phase diagram of a dilute TIM in the $\Gamma-p$ plane. Senthil
and Sachdev  
 \ct{senthil96} studied
 the quantum  transition along the vertical line which terminates
at the multi-critical point ($p_c, \Gamma_o$). 
In the long-range connected case, the phase diagram is qualitatively similar.  However, 
$p_c$ depends on $\sigma$. 
} 
\end{figure}

The experimentally 
studied systems ${\rm
 LiHoF_4}$ which
are suitably  modeled by transverse Ising spin glass models however are
complicated because of the presence of long-range dipolar interactions
 \ct{bitko96}. Moreover
in a metallic system, the presence of long-range RKKY interaction
is expected to modify the critical behaviour \ct{dobro05}.
Question
therefore remains whether a non-trivial long-range interaction modifies the 
exponents associated with an IRFP. 
 To address this particular issue, a quantum Ising spin-glass
model with long-range random interactions, where interaction between any two
spins separated by a distance $r_{ij}$  is $J_{ij}/r_{ij}^{\alpha}$ with
random $J_{ij}$'s,
 was introduced \ct{dutta02}. 
The perturbative  RG calculations \ct{read95} of the above spin-glass 
model around the upper critical
dimension failed to locate any stable weak coupling fixed point 
 in the non-mean-field region and a  RG flow towards the strong coupling
was observed. This immediately points to the existence of 
an IRFP for the above spin-glass\ct{dutta02} . However, 
no quantitative information can be derived
regarding the nature IRFP in the above spin-glass model.

In this communication,  we study the quantum transitions in a 
novel TIM with ferromagnetic interactions defined on a long-range bond dilute
lattice. It should be emphasized at the outset that in this model there is
 no long-range random interaction, nevertheless, there are
 long-range connection probabilities which effectively  
simulate
a long-range interaction that non-trivially affects the exponents at IRFP.
 The Hamiltonian of the model is 

\be
H = - \sum_{(i,j)} J_{ij} S_i^z S_j^z - \Gamma \sum_i S_i^x
\ee
where $S_i$'s are Pauli spin matrices and $\Gamma$ is a non-commuting
 transverse field.
The bond $J_{ij}$ is chosen from the binary distribution
\be
P(J_{ij}) = \frac {p}{r_{ij}^{d+\sigma}} \delta (J_{ij} -J) 
+ \left(1 - \frac {p}{r_{ij}^{d+\sigma}} \right) \delta(J_{ij})
\ee
where $p$  ($0 < p \leq 1$) denotes the connection probability between
 the nearest
neighbour sites, $d$ is the spatial dimension and the  parameter 
$\sigma$ determines the range of connection. The parameter $\sigma$
is restricted to non-zero positive values to ensure extensivity.
The Hamiltonian refers to a pure TIM which resides on a lattice where 
any two sites separated by a distance $r$ is connected with a ferromagnetic
bond of strength $J$ with a probability  falling algebraically
with the distance between the sites as $p/r_{ij}^{d+\sigma}$. 
Clearly, for smaller values of
 $\sigma$ long-range connections are dominant. 
    
An infinite randomness critical fixed point is characterised by three exponents
\ct {fisher92}: At the critical point, the smallest
 energy gap $\Delta E$ is related to the linear
dimension $L$ of the {\it rare region} as $-\ln\Delta E \sim L^{\psi}$ which
is a signature of activated quantum dynamical scaling. 
Analogously, the magnetic moment of the cluster grows as $L^{\phi \psi}
$. The typical size of such a cluster (the correlation length) diverges as
$\xi \sim \delta^{-\nu}$ where $\delta$ measures the deviation from the
quantum critical point. All the bulk exponents are
expressed in terms of the exponents $\phi, \psi $ and $\nu$
 \ct{fisher92, fisher99}. 
Senthil and Sachdev \ct {senthil96} explored the  phase transition
in the same TIM described by the Hamiltonian (1)
on a conventional dilute lattice where only the nearest neighbour sites are 
connected with 
a ferromagnetic bond of strength $J$ with probability $p$ so that 

\be
P(J_{ij}) = {p} \delta (J_{ij} -J) 
+ (1-p) \delta(J_{ij}),
\ee
where $(i,j)$'s are nearest neighbour sites.  
 The zero-temperature phase diagram
of the model in the $\Gamma-p$ plane is shown in the Fig.~1 \ct{harris74}.  
For $\Gamma=0$, the
system undergoes a geometrical phase transition at the percolation threshold
$p_c$. For $p>p_c$, at a  sufficiently large $\Gamma = \Gamma_c(p)$,  
the quantum fluctuations destroy the long-range
order. However, even at $p_c$ there exists an infinitely 
connected cluster with fractal dimensions $d_f (< d)$.  Even for
 a one-dimensional TIM, the critical transverse field is finite and thus
 a finite 
amount of transverse field is required to destroy the long-range order at $p=p_c$. This
leads to the existence of 
a multi-critical point with $\Gamma = \Gamma_o$ and $p=p_c$ as pointed out
 by Harris \ct{harris74}. Senthil and
Sachdev studied the transition along the vertical line ($p=p_c$ and $\Gamma
< \Gamma_o$) where the quantum transition is dictated by the percolation
fixed point $p_c$.   
The quantum critical exponents are solely determined by  the classical percolation 
exponents.

To comprehend the nature of  IRFP associated with the quantum transition 
of the TIM defined through Eqs.~(1) and (2), it is necessary
 to explore the associated
percolation transition  
with
long-range connection probabilities\ct{schulman83}. To study  
the critical behaviour of the above percolation model, the mapping
of the same to the corresponding $q$-state Potts Model 
(in the limit $q \to 1$)
\ct{kasteleyn69}
 turns out to be useful. The resulting Hamiltonian is 

\be 
H = - \sum_{ij} \frac {K}{r_{ij}^{d+\sigma}} \left(\prod_{\alpha=1}^{q} 
\delta_{{S_i}^{\alpha}, S_j^{\alpha}} -1 \right)
\ee
where $S_i^{\alpha}$'s are the $q$-state Potts model variables and
 the interaction $K$ is a function of nearest neighbour connection probability
$p$. 
Here, the 
connection probability between two sites  gets 
translated into the interaction between the spins as we have
only retained 
most dominant (or  relevant) long-range interacting
term in (4).

Long-range interacting ferromagnetic  Potts Model described by 
 Hamiltonian (4), exhibits
a non-trivial order-disorder transition  
at a finite temperature $ T_c(\sigma)$ (even when the 
spatial dimensionality $d=1$
if the range parameter $\sigma < 1$ \ct{schulman83, aizenman88}). 
The marginal case
$d=\sigma=1$ shows a topological transition as in the inverse-square 
Ising model \cite{anderson71,dutta03}. In the corresponding 
percolation model (with
connection probability $p/r_{ij}^{d+\sigma}$) therefore, an infinite 
percolation
cluster always exists if the nearest neighbour connection probability
$p$ is greater than $p_c(\sigma)$
and there is a percolation transition 
at  $p = p_c(\sigma)$. For $p<p_c(\sigma)$, infinitely connected cluster
disappears. The percolation threshold  $p_c(\sigma)$ 
 shifts to
the lower value of $p$ as the range parameter $\sigma$ decreases, $i.e.$,
the connection probability between two distant sites gets lower. 
At $p_c(\sigma)$, however, there
exists an infinite cluster with fractal dimension $d_f (d>d_f>1)$.  
Using the same argument as given above, we conclude that the phase diagram
of the TIM on a long-range connected lattice is qualitatively similar 
to that of the conventional dilute magnet \ct{harris74} (Fig.~1). 
A similar phase diagram
is also expected for the long-range connected TIM even for $d=1$ if
 $\sigma <1$ \ct{dutta03}. 
  
In the same spirit as in
reference [13], we shall now examine the nature of transition
at the percolation threshold $p_c (\sigma)$ with $\Gamma < \Gamma_o$
(Fig.~1).
 Along this vertical line, the disorder averaged static correlation 
function between
two spins separated by a distance $x$ ( which is actually the probability that
two sites separated by a distance $x$ belongs to the same cluster) is given as
\ct{stauffer92}
${\overline C(x)} \sim  1/ { (x^{(d-2 +\eta_{LR} (\sigma))}})$   
where $\eta_{LR} (\sigma)$ is the anomalous dimension exponent 
of the percolation
transition with long-range connections. This 
algebraic fall shows that the 
 vertical 
line is critical with transitions  dictated by the percolation
fixed point $p_c(\sigma)$.   

To derive
the percolation exponents, let us use  the Potts model
Hamiltonian (4) in the $q \to 1$ limit (where the order disorder transition 
in the Potts model is always continuous).  In the continuum limit, the 
transition  in the above Potts model is described in 
terms of   
a cubic model  with a long-range interaction term. The corresponding 
Landau-Ginzburg-Wilson action is
\ct{priest76}

\ba
{\cal H} &=&  -\frac 1 {2} \int(r + ak^2 + bk^{\sigma}) 
\sum_{i=1}^q Q_{ii}(k)Q_{ii}(k) \nonumber\\
&+& w \int \int \int  \sum_i Q_{ii}(k_1)Q_{ii}(k_2)Q_{ii}(k_3) \delta(k_1 + k_2
+k_3) 
\ea
where $Q_{ii}$ is a diagonal traceless  $q \times q$ tensor and $\int$ denotes
the integration over $\vec k$ in the first Brillioun Zone, $r$ and $u$ are the
mass and the coupling terms respectively. 
The term $bk^{\sigma}$ arises due to
the long-range interaction in the Hamiltonian (4). At the Gaussian
level this term dominates over the short-range term  $ak^2$
as long as  $\sigma <2$. For higher values of 
$\sigma$ ( $\geq 2$), long-range interactions is irrelevant and 
therefore exponents
of the  Potts model are of short-range nature.  
In higher-orders when the effect of the coupling term $w$ is perturbatively
 included, 
the short-range term $ak^2$ picks up the anomalous dimension
$\eta_{SR}$.  We therefore conclude that the long-range interaction 
is relevant as long as $\sigma <2 - \eta_{SR}$. The most interesting 
feature of any long-range interacting 
spin model, described by action, (5) is that the
 term $bk^{\sigma}$ does not get renormalised at
any order of perturbation and hence the anomalous dimension
for the long-range interacting system sticks to the value 
$\eta_{LR} = 2 - \sigma$ for all values of $\sigma < 2 - \eta_{SR}$ \ct{fisher72}. It can 
also be checked by
simple dimensional counting of different parameters of action (5) that 
the upper-critical
dimension and  the range follow the relation
 $d_u = 3 \sigma_u$, $i.e., $ for a
given $d$ if the range parameter $\sigma < \sigma_u (=d/3)$, the interactions
are long-ranged enough to render mean field theory to be exact. 
For $\sigma =2$ (short-range case), the upper critical dimension is as 
usual $6$.

Let us now restate the above results for the equivalent long-range percolation
problem. As expected, the exponents depend on $\sigma$. In particular,  
the anomalous dimension exponent, $\eta_p(\sigma)$, of a
 percolation transition is related
to the fractal dimension of the lattice at $p_c$ through the scaling
relation \ct{stauffer92}  $d_f(\sigma) = (d+2 -\eta_p(\sigma))$. In the 
 long-range model,
$\eta_p (\sigma) = \eta_{LR}(\sigma) = 2 - \sigma$ which immediately leads to
the result $d_f(\sigma) = (d+\sigma)/2$. The fractal dimension 
at the percolation threshold
is therefore exactly known for long-range percolation problem
for all values of $d$ and $\sigma$. The other exponents of the percolation
transition can be  obtained by perturbative calculation around the upper
critical dimension \ct{priest76}
 $e.g.,$ to the first order in 
$\epsilon (= 3 \sigma
-d )$, the exponent $\nu_p = 1/{\sigma} + \epsilon/4 \sigma^2$. The exponents
not only depend on the range parameter $\sigma$ but also
cross over 
to the short-range value as $\sigma$ approaches $2 - \eta_{SR}$ when
the long-range interaction term $b k^{\sigma}$ becomes irrelevant. 
In the following, we shall use the standard percolation 
scaling relation for the probability that a site falls in
a finite cluster of size $s$ given as \ct{stauffer92}
\be
P(s,p) = s^{1-\tau} f(s(p-p_c(\sigma))^{1/\sigma})
\ee
where $\tau = (d+d_f)/d_f$ and $\sigma = \nu d_f$.

We are now in a position to obtain  the critical exponents related to the
quantum transitions  at $p=p_c$ and $\Gamma < \Gamma_0$.
 Let us first look at the exponent $\psi$. The
energy gap $\Delta E$ at the percolation threshold of a cluster of size $L$
scales as $\Delta E \sim \exp (-c L^{d_f})$ where $c$ is a non-universal
constant.   
The logarithmic scaling of the correlation length
with the frequency can be easily  verified also by calculating the dynamical
response function using the percolation scaling relation (6) \ct{senthil96}.
The exponent $\psi$ is therefore given as $\psi = d_f = (d+ \sigma)/2$. 
The important 
features which constitute the essential theme of this communication
 need to be highlighted 
here: i) the exponent $\psi (\sigma)$ is exactly known for all values 
of $\sigma$
and $d$ while in the conventional case the exponent is exactly known only in
the spatial dimension \ct{stauffer92}, $d=2$ ii)  $\psi$
depends on the range of interaction $\sigma$ and thus gets modified 
as the range parameter $\sigma$ is tuned and iii) $\psi (\sigma)$ undergoes 
a crossover to the short-range value when $\sigma \geq  2 - \eta_{SR}$.
To evaluate the exponent $\phi$, let us now use the scaling relation of 
magnetisation in a weak magnetic field $h$ at the quantum critical point
 given as \ct{fisher92}
\be
m(h) \sim \left[\ln(\frac {h_o}{h}) \right]^{\phi - d/\psi}
\ee
where $h_o$ is a non-universal constant. In the present TIM, the magnetisation
at $p=p_c$ scales as
\be 
m(h) \sim \left[\ln(\frac {h_o}{h}) \right]^{2 - (d+d_f)/d_f}
\ee
Comparing Eqs.~(7) and (8) and using $\psi = d_f$ and 
 the percolation exponents \ct{stauffer92}
 $\beta_p  = (\tau -2)/\sigma$, $\gamma_p =
(3-\tau)/\sigma$,
 we  have
$\phi = (d - \beta_p/\nu_p)/d_f =1$.

As in the conventional dilute magnet \ct{senthil96}, in the
 present long-range case also
both the sides 
of the transition point $p_c$  are found to be flanked with Griffiths phase
 with 
continually varying exponents. For example, in the disordered phase ($p<p_c(\sigma)$), the disorder averaged
 imaginary part of local dynamical
susceptibility goes as

\be
\overline {{{\chi_L''(\omega)}}}\sim (\omega)^{d/z}-1
\ee
where in arriving at Eq.~(9), we have used the same functional form of 
the percolation scaling function 
$f(x)$ (Eq.~6) as the conventional case \ct {stauffer92} with exponents depending on $\sigma$. 
Eq.~(9) shows that the paramagnetic phase is gapless with a power-law density
of states at low energy. This  power-law singularity leads to off-critical
GM singularities, $e.g.$, the average local susceptibility diverges even
in the paramagnetic phase as $T^{d/z-1}$ when the temperature
 $T \to 0$ \ct{vojta06}.
The dynamical exponent $z$ is related to the
percolation correlation length  $\xi_p$
as 
$z = \xi_{p}^{d_f} = (p_c-p)^{-\nu d_f}$. The dynamical exponent
varies continuously with $(p_c-p)$ and diverges at the percolation threshold.
We immediately 
conclude that
 the dynamical exponent $z$  
increases with increasing $\sigma$ leading to
a stronger divergence of $z$ at $p_c$, indicating an  enhancement in
 Griffiths-McCoy singularities.  
 Therefore, 
the strength of the Griffiths phase also varies as $\sigma$ is 
tuned.

In conclusion, the transitions in  a TIM  at the percolation threshold 
of  a dilute lattice
 with long-range connection probabilities provide an ideal situation where
the effect of long-range connections on the IRFP can be extensively
studied.  
The exponents $\psi$ and $\nu$ of IRFP are found to 
depend non-trivially on the 
range parameter $\sigma$ and cross over to the short-range values when
the range parameter $\sigma \geq  2 - \eta_{SR}$. However, for lower values
of $\sigma (< \sigma_u)$, when the long-range connections are prominent,
the exponents are described in terms of mean field percolation
exponents \ct{stauffer92}. Most importantly, the exponent
$\psi$ is exactly known for all values of $d$ and $\sigma$  and increases
as $\sigma$ increases.
 The strength of GM singularities also get enhanced
for higher values of $\sigma$. This is expected because  for smaller values of
 $\sigma (<\sigma_u$)  the model   becomes more mean field like 
and 
 GM singularities should be weak.
A recent numerical study (based on the 
finite size scaling argument) 
 of a percolation
transition in a power-law dilute chain \ct{albuquerque05} in $d=1$ suggests that
 the scaling relation
$d_f = (d +\sigma)/2$, obtained from renormalisation group arguments,  may 
not be valid when
long-range connection probabilities are dominant,
 $i.e.$, for $0 < \sigma \leq 1/2$. However, for higher values
of $\sigma$
the numerical results corroborate the renormalisation
group results. No such numerical study is available for the percolation
transition with power-law dilution in dimensionality greater than unity. 
Question therefore remains whether one should use the value of
 $d_f$ as obtained
from the Potts model analogy in smaller $\sigma$ region for $d=1$. 
Also, Vojta and Schmalian \ct {vojta05}
showed that for dilute quantum rotors 
 the quantum dynamics
is  of conventional (power-law) nature  
(for transitions below the multicritcal point)
 with the dynamical exponent $z =d_f$ 
and the correlation
length exponent $\nu = \nu_p$. The scaling relation employed in Ref.~
[14]  should in principle also work in the case of the present
 dilute lattice with long-range connections. We should also mention
that the present work may also be relevant for the studies of quantum
phase transitions on some interesting network models \ct{moukarzel02}.

The authors gratefully acknowledge interesting discussions with 
S. M. Bhattacharjee, D. Chowdhury, C. Dasgupta, R. Narayanan, T. V.
Ramakrishnan and V. Subrahmanyam. AD acknowledges Uma Divakaran for her help.

\end{document}